# BoostJet: Towards Combining Statistical Aggregates with Neural Embeddings for Recommendations

Rhicheek Patra
EPFL
rhicheek.patra@epfl.ch

Egor Samosvat
Yandex
sameg@yandex-team.ru

Michael Roizner
Yandex
roizner@yandex-team.ru

Andrei Mishchenko
Yandex
druxa@yandex-team.ru

## ABSTRACT

Recommenders have become widely popular in recent years because of their broader applicability in many e-commerce applications. These applications rely on recommenders for generating advertisements for various *offers* or providing content recommendations. However, the quality of the generated recommendations depends on user features (like *demography*, *temporality*), offer features (like *popularity*, *price*), and user-offer features (like implicit or explicit *feedback*). Current state-of-the-art recommenders do not explore such diverse features *concurrently* while generating the recommendations.

In this paper, we first introduce the notion of Trackers which enables us to capture the above-mentioned features and thus incorporate users' online behaviour through statistical aggregates of different features (demography, temporality, popularity, price). We also show how to capture offer-to-offer relations, based on their consumption sequence, leveraging *neural embeddings* for offers in our Offer2Vec algorithm. We then introduce BoostJet, a novel recommender which integrates the Trackers along with the neural embeddings using MatrixNet, an efficient distributed implementation of gradient boosted decision tree, to improve the recommendation quality significantly. We provide an in-depth evaluation of BoostJet on Yandex's dataset, collecting online behaviour from tens of millions of online users, to demonstrate the practicality of BoostJet in terms of recommendation quality as well as scalability.

## 1 INTRODUCTION

The rapid growth of Internet has affected everyone's daily life. People spend more and more time on the internet with different applications for multiple purposes like searching or shopping. Fueled by such online applications, recommendations have emerged as a popular paradigm to enable users to quickly navigate through the huge volume of data available online. Recommenders try to *learn* about the preferences of the users by observing their rating history. Recommenders mainly leverage *content-based* or *collaborative* filtering techniques. Content-based filtering techniques are typically based on offer features (e.g., features for a movie could be specific actors, directors, genres, plot, etc) as well as the user features which are aggregates of the features specific to the offers preferred by a given user. Collaborative filtering techniques are based on collecting and analyzing a large amount of information regarding users' behaviors, activities or preferences and predicting the future preferences of users based on their similarity to other users. Another crucial observation in these online platforms is that users' interests and the popularity of offers shift very fast, which poses great technical challenges to existing recommender systems for recommending the right offers at the right time.

### 1.1 Motivation

We now highlight the most crucial factors which are essential for improving the recommendation quality. There is a lack of research work exploring such diverse factors concurrently while generating the recommendations.

**Content.** Content-based factors capture how different offer attributes affect the users' future preferences. To get an intuition, consider the case of advertisements for which these attributes could be popularity of the shop among the previous customers for any given *offer brand* like Apple or *offer category* like iPhone. There have been previous work on designing efficient content-based systems [21] based on such offer attributes.

**Temporality.** Besides the problem of information overload for online services, the interests of users also vary drastically with time. Incorporating such time drifting preferences leads to technical challenges. To get an intuition about these challenges, consider the emergence of new products which might change the preference of customers. Such preference drift can be related to seasonal changes or specific holidays, which might lead to characteristic shopping patterns. A change in the family structure can drastically change shopping patterns which is another such example of preference drift, driven by localized factors. Likewise, individuals gradually change their taste in movies and music. There have been also several work on incorporating such temporality in recommendations [16, 6, 12, 9].

**Demography.** Demographic information is essential for making relevant recommendations to users. Demographic information, sometimes called the *target audience*, of a product is a collection of the characteristics of the customers who are interested in that specific product. Recently, the recommendation quality has been improved by designing recommenders which leverage the features derived from both the product demographics as well as the demographic characteristics extracted from the users [26].

**Price.** Current rating-based approaches ignore monetary factors which are crucial to the shop's revenue as well as the users' product adoption decisions, namely the price and whether such price would be considered acceptable by an individual. For e.g., it would be irrelevant for a recommender to provide a high-priced special edition of a highly relevant product if its price is significantly outside the purchasing range of the target user. Naturally, there exists inherent uncertainty in users' purchase decisions due to people's innate valuation [15]. There has been some research devoted to designing efficient recommendations to incorporate monetary factors [20].

### 1.2 Overview

In this paper, we introduce the notion of TRACKERS which enables us to incorporate these factors concurrently. Then, we present a novel recommender BoostJet which combines these TRACKERS along with our proposed Offer2Vec's neural offer embeddings using MatrixNet [13][1] which is an optimized implementation of gradient boosting machines (GBM) [7].

BoostJet computes the recommendations as follows. First, BoostJet generates the TRACKERS which are statistical aggregates of users' activity capturing factors of different types (content, temporal, demographic, or monetary). Second, BoostJet generates the offer embeddings to capture the higher-dimensional relation between different offers in a given shop based on their consumption order by different users. These embeddings are generated using the proposed Offer2Vec algorithm, our modification of Doc2Vec [18], by treating each *user session*, in a given shop, as a *document* and offers in this session as *words*. Finally, with the help of MatrixNet we combine these features by posing the recommendation task as a classification problem in BoostJet.

Another important feature of BoostJet is the scalability which is crucial for deploying such a recommender, incorporating multiple different recommendation factors, in a practical large-scale scenario involving millions of users and items.

### 1.3 Contribution

The main contributions of this paper are summarized as follows.

(1) We introduce the notion of TRACKERS to incorporate different recommendation features namely content, temporal, demographic, or monetary. We also capture the offer-to-offer relations based on the consumption patterns of different users using Offer2Vec. These features are generated by either computing statistical aggregates of the respective consumption events or measuring the similarity between any given offer and the target user's recent consumption session.

(2) We present a novel recommender BoostJet which combines these generated features (TRACKERS and Offer2Vec) using gradient boosted decision trees (MatrixNet). We empirically show that BoostJet significantly improves the recommendation quality provided by i-Tencent [14], an implicit recommendation algorithm employed in practice by Tencent Inc.

(3) We provide an in-depth empirical evaluation of BoostJet in terms of recommendation quality and scalability. We deploy

---

[1]An open-source version of MatrixNet library is available at https://github.com/catboost/catboost.

BoostJet on Yandex's dataset, with tens of millions of users and offers, to evaluate how it performs in practice.

### 1.4 Roadmap

We discuss about the features (TRACKERS and Offer2Vec) used in BoostJet along with the necessary steps for generating them in § 2. In § 3, we present a brief overview of MatrixNet, followed by the feature aggregation, training, and recommendation generation (from selected candidates) in our novel recommender BoostJet. We then provide our detailed empirical evaluation of BoostJet in terms of recommendation quality and scalability in § 4. We provide the related work in § 5 before concluding the paper in § 6.

## 2 BOOSTJET FEATURES

In this section, we present the features that are used in BoostJet for computing the recommendations.

### 2.1 TRACKERS

The underlying idea behind our TRACKER-based featurization is simple: *the TRACKERS compute the statistical aggregates of user activities for different factors (content, temporal, demographic, or monetary).* Then, these TRACKERS are used as input to MatrixNet for determining the preferential scores of users towards offers. TRACKERS could be grouped into various classes, based on the factors they are tracking, which are provided as follows.

**Content TRACKERS.** These trackers compute the statistical aggregates based on the content-based features of the offers. Table 1 presents the basic content-based feature information which are used in these TRACKERS. These TRACKERS also compute aggregates for combination of different content features. For example, the number of clicks by a user in a given shop is aggregated by CLICKS [ShopID, UserID] whereas the number of purchases for a given brand, corresponding to an offer, in a given shop is aggregated by PURCHASES [ShopID, OfferBrand].

Content TRACKERS can also compute normalized statistical aggregates. A sample TRACKER, computing such normalized aggregate, is $\frac{\text{PURCHASES[ShopID, OfferID, OfferBrand]}}{\text{PURCHASES[ShopID, OfferBrand]}}$ that computes the fraction of the number of purchases for a given offer among all the offers in the same shop with the same brand as the the given offer. For e.g., this TRACKER can compute the fraction of the number of iPhone-7 smart-phones which are sold among all Apple products from eBay.

**Temporal TRACKERS.** These TRACKERS capture the temporal patterns among users or offers. As we mentioned before, seasonal changes, specific holidays, or users' behavioural changes could lead to characteristic shopping patterns. Table 2 demonstrates the different types of temporal TRACKERS used. To get an intuition, consider CLICKS [ShopID,UserID,SincePrevTime] which computes the time difference since the previous time the given user performed click action in this given shop whereas ADDS [ShopID,OfferID,LastWeek] tracks the number of add actions for the given offer in the given shop during the last week.

The temporal TRACKERS also compute normalized aggregates to capture the time-based trends of offers or users. For e.g., the fraction of the number of any actions for the given offer in the given shop on the last day among those received in the last week can be tracked using $\frac{\text{ANY[ShopID,OfferID,LastDay]}}{\text{ANY[ShopID,OfferID,LastWeek]}}$. This type of TRACKERS captures the trending offers in a given time window (day, week or month) among users from a given shop.

| Notations | |
|---|---|
| OfferName | Name of an offer. |
| OfferNameCats | Automatic categories based on an offer's name. |
| OfferBrand | Brand name of an offer. |
| MarketModel | Model-id of the given offer in market (Yandex's shop aggregator). |
| MarketCategory | Manual category of an offer in the market. |
| MarketVendor | Vendor's unique identifier in the market. |
| ShopId | Unique identifier of a shop. |
| UserId | Unique identifier of a user. |
| OfferId | Unique identifier of an offer. |
| Action | User behaviour which could be *click*, *detail*, *add*, or *purchase*. We denote *any* action for any of these four possible primary actions. |

Table 1: Basic content information for TRACKERS.

| Notations | |
|---|---|
| LastDay | Tracks the number of actions in the last day. |
| LastWeek | Tracks the number of actions in the last week. |
| LastMonth | Tracks the number of actions in the last month. |
| SinceFirstTime | Tracks the time difference since the first occurrence of the event. |
| SincePrevTime | Tracks the time difference since the previous occurrence of the event. |

Table 2: Types of temporal TRACKERS.

**Demographic TRACKERS.** These TRACKERS compute the statistical aggregates based on the demographic information of users. As we mentioned before, the *target audience* for any given offer is related to the demographic information of the users who are interested in the offer. However, for our dataset, there are hundreds of demographic region, identified by their region-ids, based on users' demographic information. We use a binning strategy to cluster the regions into bins. Based on the demographic information in the dataset and the frequency of these regions, we partition the regions into three bins: *Moscow*, *Saint-Petersburg*, and *Others*. To get an intuition regarding the functionality of demographic TRACKERS, consider CLICKS [ShopID,OfferID,RegionID] which tracks the number of clicks from a specific demographic region corresponding to any given offer from a given shop.

We can also have normalized demographic TRACKERS following the same technique as earlier ones. A sample example of such normalized demographic TRACKER is $\frac{\text{ANY[ShopID,OfferID,RegionID]}}{\text{ANY[ShopID,RegionID]}}$ which captures the fraction of any actions from users in a given region corresponding to a given offer.

**Price TRACKERS.** These TRACKERS compute the statistical aggregates based on the price information of offers as well as the monetary features of users like price range. Similar to demographic TRACKERS, we use a binning technique which partitions the price range into $M$ [2] logarithmic bins depending on the maximum and minimum price of the products in our dataset. Hence, a price tag of $L$ is mapped to a bin with id $\lfloor \log_{10}(L) \rfloor$. For example, ANY [ShopID,PriceBin] tracks the number of any actions for different price bins in a given shop. Hence, this type of TRACKERS can capture the trending offers based on the price tags associated with the corresponding offers. Similar to previous TRACKERS, we can also compute normalized price TRACKERS.

**Hybrid TRACKERS.** The above-mentioned TRACKERS could also be combined to capture more complex recommendation features. For example, PURCHASES [ShopID,PriceBin,RegionID,LastWeek] tracks the number of purchases in the last week in a given shop for different price bins from users belonging to different demographic locations. Moreover, similar to the earlier TRACKERS, we can also compute normalized hybrid TRACKERS.

**Computation.** TRACKERS are computed for a particular user-offer pair as soon as all content information (Tables 1 and 2) is defined. However to make this computation feasible, we first pre-aggregate the users' interaction history, and then calculate all the possible TRACKERS which we might need later. We perform this pre-aggregation effectively using Map-Reduce operations. More precisely, we compute all the above-mentioned TRACKERS using one standard Map-Reduce operation which is as follows.

(1) *Map phase.* BOOSTJET emits user's history record with a key defined by the TRACKER.
(2) *Reduce phase.* BOOSTJET computes the aggregation function defined by the TRACKER.

We next use the stored pre-aggregated statistics during BOOSTJET's train pool generation (see § 3.2). For this purpose, we just need to lookup for particular TRACKERS' values corresponding to a user-offer pair in the query.

### 2.2 OFFER2VEC

We now describe our OFFER2VEC method which enables us to generate embeddings for offers in a latent space and capture the offer-to-offer relations based on the consumption patterns of these offers by different users.

**Doc2VEC overview.** Most state-of-the-art machine learning algorithms require the input to be represented as a fixed-length feature vector. In the domain of text-based machine learning models, one of the most common fixed-length features based model is bag-of-words. Despite their popularity, bag-of-words models suffer from two crucial drawbacks. First, they lose the ordering of the words. Second, they ignore the semantics of the words. For example, "powerful", "strong" and "Paris" are equally distant. To address this issue of incorporating the order in the model, Mikolov et al. proposed DOC2VEC [18], an unsupervised algorithm that learns fixed-length feature representations from variable-length pieces of texts, such as sentences, paragraphs, and documents. Two training models were proposed for DOC2VEC. The first one is a Distributed Memory model (DM) which incorporates the word ordering, and the second one is a Distributed Bag of Words model (DBOW) which ignores the word ordering in the texts. Furthermore, the DM model considers the concatenation of the document vector with the word vectors to predict the next word in a text window whereas the DBOW model ignores the context words in the input, but forces the model to predict words randomly sampled from the document in the output. In this paper, we focus on the DM model as it is shown to perform better in practice.

---
[2] In our experiments, we have $M = 7$.

*Distributed Memory Model.* Every word is mapped to a unique vector, represented by a column in a matrix $W$, and every document is mapped to a unique vector, represented by a column in matrix $D$. More formally, given a sequence of $T$ training words' vectors $w_1$, $w_2$, $w_3$, ..., $w_T$, a maximum hop size of $k$, and a document vector $d$, the objective of the distributed memory model is to maximize the average log probability

$$\frac{1}{T} \sum_{t=k}^{T-k} \log P(w_t | w_{t-k}, ...., w_{t-1}, w_{t+1}, ...., w_{t+k}, d) \quad (1)$$

Doc2Vec is trained using stochastic gradient descent via back-propagation and the log probabilities (Equation 1) are computed using softmax. For optimization purpose, Doc2Vec uses an efficient approximation for softmax, namely, *hierarchical softmax* which is a binary Huffman tree where short codes are assigned to frequent words. Doc2Vec uses *negative sampling* to distinguish the target word $w_t$ from negative samples drawn from some noise distribution and also improves the training time significantly by updating the weights corresponding to only the target word along with the sampled ones (instead of the whole vocabulary).

**Offer2Vec in BoostJet.** Offer2Vec generates offer embeddings (vectors) which are leveraged by BoostJet to generate features capturing the consumption pattern. A user's profile consists of the set of offers that the user interacted with. Any such user profile is partitioned into $\delta$-distant sessions ($S$). We define the notion of a session in a user profile as follows.

*Definition 2.1 ($\delta$-DISTANT SESSIONS IN A USER PROFILE).* Let $o_k$ and $o_{k+1}$ denote two consecutive offers in the profile of a user $u$, with interaction timestamps $t_{u,o_k}$ and $t_{u,o_{k+1}}$, such that $t_{u,o_k} \leq t_{u,o_{k+1}}$. Given $o_k$ belongs to a session $S_u^l$, the offer $o_{k+1}$ is added to $S_u^l$ if $t_{u,o_{k+1}} \leq t_{u,o_k} + \delta$. Otherwise $o_{k+1}$ is added as the first offer in a new session $S_u^{l+1}$.

Each of these $\delta$-distant sessions are treated as a separate document for the Offer2Vec model. We next train the Offer2Vec model on these $\delta$-distant sessions from all the users. The Offer2Vec model then outputs an $n$-dimensional vector for each of the offers present in the database (vocabulary). The offer vectors, provided by Offer2Vec, enable us to generate vectors for any given session of a user. We compute the vector for any user's most recent session as the mean of all the vectors corresponding to offers in that session. Finally, the pattern feature is generated by computing the cosine similarity between the vector corresponding to the most recent session of a specific user and any given offer to indicate the likelihood of this offer to be preferred next.

## 3 BOOSTJET

In this section, we present BoostJet which combines the different recommendation features (Trackers and Offer2Vec) using MatrixNet. We first provide a brief overview of MatrixNet along with its optimizations. We then discuss how we leverage the generated Trackers and Offer2Vec features for training BoostJet. We next describe our candidate generation technique which selects a set of candidate offers for a given user. Finally, we show how BoostJet provides the most relevant recommendations for a given user by applying the classification task over the generated set of candidate offers.

### 3.1 MatrixNet

MatrixNet [13] is an optimized implementation of Gradient Boosting Machine (GBM) [7] with stochastic gradient boosting [8]. We denote the $n$-dimensional input feature by $x \in \mathcal{R}^n$, the target label by $y \in \mathcal{R}$, and the evaluation function over the input feature by $F(x) \in \mathcal{R}$. Given an appropriate differentiable loss function $L(y, F(x))$, MatrixNet could be leveraged to solve multiple machine learning tasks like classification or regression. The underlying idea is to formulate the evaluation function $F(x)$ iteratively as follows.

$$F_m(x) = F_{m-1}(x) + \gamma h_m(x)$$

At each step, $h_m(x)$ is selected from some class of weak learners in order to fit the pseudo-residuals defined as follows.

$$r_{im} = -\partial L(y_i, F_{m-1}(x_i)) / \partial F_{m-1}(x_i)$$

This fitting step can be interpreted as modifying our training set to $\{x_i, r_{im}\}_{i=1}^n$. In this way, we choose $h_m(x)$ such that adding it to the sum would most rapidly minimize the loss.

The learning rate in MatrixNet is controlled by the "shrinkage" parameter $\gamma$. Ideally, a lower value of $\gamma$ leads to a more precise solution but with higher number of iterations ($M$). Furthermore, the error on the training set reduces with more iterations ($M$) but might lead to over-fitting with very high number of iterations. Thus, $\gamma$ and $M$ are the main parameters which control MatrixNet's performance.

After the training phase, MatrixNet provides a summary of feature importance (effect) which highlights the strength of each feature in terms of how often it was used in the boosting iterations.

We apply MatrixNet for a binary classification task and define our loss function with negative log likely-hood as follows.

$$L(y_i, F_m(x_i)) = -y_i \log(P_m(x_i)) - (1 - y_i) \log(1 - P_m(x_i))$$

where $P_m(x_i)$ is the predicted probability of $y_i = 1$ computed using a logistic transformation of $F_m(x)$ as follows.

$$P_m(x_i) = (1 + \exp^{-1}(F_m(x_i)))^{-1}$$

Compared to TreeNet [3], another implementation of GBM, MatrixNet uses oblivious decision trees as the set of weak learners. An oblivious decision tree tests the same feature for all nodes at the same depth. More precisely, the nodes at the same depth in MatrixNet share not only a feature but also a border. Typically, a decision tree in MatrixNet has 6 different splits and $2^6$ leaves. Such trees can be interpreted as a $p$-dimensional *matrix* indexed by a $p$-dimensional boolean variable, where $p$ is the number of different splits in the tree. The choice of weak learners and some regularization heuristics led to MatrixNet's exceptional performance in practice. It is widely used for multiple applications like web search ranking, ads click prediction, bot detection [13, 25, 24].

### 3.2 Training

We now present how we train the MatrixNet model using these generated features.

---

[3]https://www.salford-systems.com/products/treenet

**Time window.** The goal of BoostJet is to recommend offers a user might like in the future. Thus, for the training purpose, we sort the training events based on the associated timestamps and then partition into two parts: the *past* and the *future*. We use the first part for feature generation whereas the second part is used for training the MatrixNet model. The timespan is demonstrated in Figure 1.

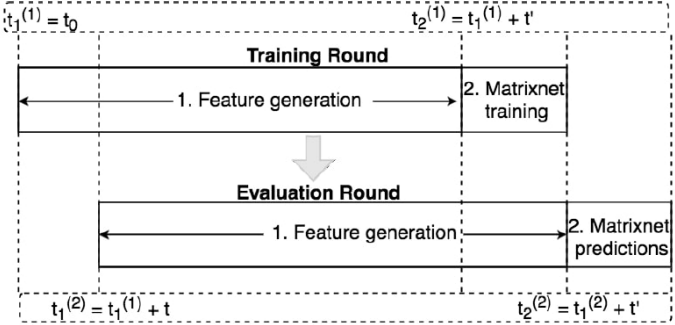

Figure 1: Training and evaluation in BoostJet.

During the evaluation round, we move the time window and recompute the features (Trackers and Offer2Vec). Then, the MatrixNet model predicts the future events using these more up-to-date features as input.

**Classification.** We approximate the recommendation problem as a classification task in BoostJet to leverage the features with MatrixNet. Given a user, a set of candidate offers, and the different recommendation features from Trackers and Offer2Vec, our goal is to predict the probability of a user clicking an offer in the future. The MatrixNet algorithm maximizes the likelihood of correct classification on the training dataset. The recommendation task is achieved by recommending the offers in a candidate set with the highest probabilities.

**Negative sampling.** All online platforms, collecting user interactions, are only able to collect positive feedback like click or purchase events. However, a classification task also needs negative samples to be trained. In BoostJet, we select recommendations from top-$N$ popular offers in the given shop (see § 3.3), that is why we choose our negative samples randomly from top-$N$ popular offers in this given shop.

Hence, our classification task learns how to classify a click and non-click from most popular offers in a given shop. More specifically, for each click (positive sample) from a user, we select $n$ negative samples at random from the set of top-$N$ popular offers in the specific shop corresponding to the clicked offer.

### 3.3 Personalized Candidate Generation

Since a shop might have several millions of offers, computing the click probabilities of a user for all of these offers is computationally intractable. Hence, we generate a set of personalized candidate offers from these offers for which we compute the click probabilities. The accuracy of BoostJet's classification task also partially depends on the quality of these candidates on which the classification task is performed. The personalized candidate generation in BoostJet is three-fold.

- *Initial set.* For this step, we select the $20K$ most popular offers in a given shop.
- *Personalized feature computation.* In this step, we compute the best personalized features for a given user. We choose the top-5 personalized features based on the feature importance (effect) provided as an output from the MatrixNet model (mentioned in § 3.1).
- *Personalized candidate selection.* Next, we use these personalized features along with the non-personalized ones as input to the MatrixNet model and then select the top $5K$ offers from the initial set of $20K$ offers. Note that all the non-personalized features can be computed once in advance as they are the same for all users. Hence, the time needed for this step is proportional to the product of the number of personalized features used and the size of the initial set.

Finally, these $5K$ offers from the given shop are the personalized candidates for the given user. We follow these three steps for all the users.

### 3.4 Recommendation

Our recommendation task is to compute the likelihood probabilities for any given user corresponding to unseen offers in a given shop and then provide the highly predicted ones as recommendations to the user. For generating these recommendations, we use the personalized candidate generation step to filter the candidate offers for a given user. We then compute the click probabilities for these candidate offers and the given user using our trained MatrixNet model, and finally recommend the top-10 ones which have the highest click probabilities.

## 4 EVALUATION

In this section, we provide a detailed empirical evaluation of the performance of BoostJet in terms of recommendation quality as well as scalability on multiple computing nodes.

### 4.1 Experimental Setup

**Experimental platform.** We perform all our experiments on a Map-Reduce infrastructure consisting of nearly $2K$ computing nodes.

**Dataset.** We evaluate BoostJet on a dataset consisting of users' online activities from across 1250 different e-commerce sites (retailers) collected over last 3 months. Overall, there are 47,929,260 events from 13,774,733 users over 2,412,929 offers from these multiple e-commerce sites along with the timestamps corresponding to the interaction events between the users and offers. Each interaction event can be classified into four categories based on the interaction as follows.

- *Click.* These interaction events denote the instances when the users clicked on some offers which they found interesting.
- *Detail.* These interaction events denote the instance when the users were interested in the details about the offers.
- *Add.* These events denote that the offers were added to the basket by the user. However, there could be instances when users remove offers from their baskets later due to various reasons like they might find something more cheaper or interesting.

- *Purchase.* These events are corresponding to the instances when the offers were finally purchased by the users.

Table 3 shows the distribution of the interaction events into these four categories. We use this dataset for generating the features using TRACKERS and OFFER2VEC, and then train the MATRIXNET model using the generated features. We also have a test dataset consisting of the activity in the last session of 184,415 users across 8 different sites among the 1250 sites from the training dataset.

| Event types | Count | Fraction |
|---|---|---|
| Click | 10,304,597 | 0.215 |
| Detail | 2,785,971 | 0.058 |
| Add | 29,888,684 | 0.624 |
| Purchase | 4,950,008 | 0.103 |

Table 3: Distribution of interaction events.

**Evaluation metrics.** We evaluate BOOSTJET along two complementary metrics: (1) the recommendation quality as perceived by the users, and (2) the scalability in terms of the reduction in computation time for BOOSTJET when increasing the number of machines in the cluster.

*Quality.* We evaluate the quality of the recommendations in terms of discounted cumulative gain (DCG). This evaluation metric quantifies the quality of recommendation while taking into account the ranking among the recommendations. We compute the DCG metric as follows.

$$\text{DCG} = \sum_{i=1}^{N} (0.85)^{i-1} * rel_i$$

where $rel_i$ is a binary variable with a value of 1 if the offer recommended at position $i$ is relevant (i.e, clicked by the user) else it is 0. $N$ denotes the number of recommendations provided to a user.

We also evaluate the quality of the trained MATRIXNET model using Log-likelihood prediction (LLP) which is defined as follows where $\widehat{P_c}$ denotes the best constant prediction, $\widehat{P}$ denotes prediction of the MATRIXNET model.

$$c_i = \begin{cases} 1 & \text{if } i \text{ is a positive sample.} \\ 0 & \text{if } i \text{ is a negative sample.} \end{cases}$$

$$\widehat{P_c} = \frac{\sum_{i=1}^{N} c_i}{N}$$

$$\text{LLP} = \frac{LL(\widehat{P}) - LL(\widehat{P_c})}{\sum_{i=1}^{N} c_i}$$

The LLP metric measures how much better are the output predictions, provided by MATRIXNET, compared to the constant predictions in terms of log-likelihood (LL). Hence, the LLP metric measures the quality of the classification task whereas the DCG metric measures the quality of the recommendation task.

*Scalability.* We measure the scalability in terms of the reduction in the computation time as well as the achieved speedup with an increasing number of computational resources. The speedup is defined in terms of the time required for the sequential execution ($T_1$) and the time required for parallel execution with $p$ machines ($T_p$). Amdahl's law models the performance of speedup ($S_p$) as: $S_p = T_1/T_p$. Due to the considerable amount of computations, a sequential execution is computationally intractable. Hence, we compare the speedup on $p$ machines with respect to a minimum of 500 machines ($T_{500}$) instead of a sequential execution ($T_1$).

**Evaluation scheme.** We use the training dataset, consisting of all the interaction events of users over the last 3 months (13 weeks), for generating the TRACKERS (first 12 weeks) as well as training the MATRIXNET model using the generated TRACKERS (last 1 week). We then generate TRACKERS for the interaction events in the test set (next 1 week) and use these as input to the trained MATRIXNET model to generate the recommendations (only new unseen offers) as mentioned in § 3. We evaluate the training of the MATRIXNET model using the LLP metric on randomly selected 10% of the data provided as input to MATRIXNET.

## 4.2 Recommendation Quality

We now evaluate the recommendation quality of BOOSTJET and compare it with a state-of-the-art recommender leveraging matrix factorization technique for implicit feedback (I-TENCENT [14]).

**Parameter tuning.** We now explain the parameter tuning for the MATRIXNET model which is crucial to obtain the best recommendation quality. The most important parameters are the number of negative examples (#neg) as well as the shrinkage parameter ($\gamma$). The negative examples help the MATRIXNET model to clearly filter the positive samples whereas the shrinkage parameter prevents overfitting of the MATRIXNET model on the training data.

*Effect of negatives.* The negative sampling technique plays a crucial part in the training of our MATRIXNET model. For each click (positive sample) from a user, we select $n$ negative samples at random from the set of top-$N$ popular offers in a given shop corresponding to the offer clicked by the user. As we observe from Figure 2(a) that the classification task can be significantly affected, in terms of DCG, by the choice of the number of negative examples. We also observe from Figure 2(b) that the MATRIXNET converges faster with an increasing number of negatives. The convergence behaviour is due to the fact that MATRIXNET can better distinguish between positive and negative examples with more negative samples per positive sample. However, the training time for MATRIXNET also increases with more negative samples. Based on these observations, we choose 25 negative examples for each positive example in our training set while training the MATRIXNET model. We fix $\gamma$ to 0.03 for these experiments.

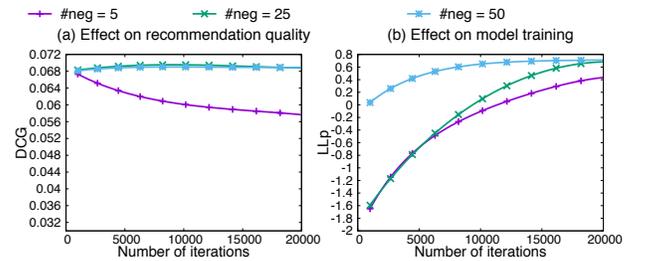

Figure 2: Effect of the number of negative examples.

*Effect of shrinkage.* We use a shrinkage parameter ($\gamma$) with the loss function $F(d)$ to reduce the overfitting of the MATRIXNET model. Figure 3 demonstrates the effect of the shrinkage parameter on the recommendation quality (Figure 3(a)) as well as on the training of the MATRIXNET model (Figure 3(b)). We observe that too high shrinkage values affect the convergence significantly due to slower convergence rate. Based on our observations, we select the value of the shrinkage parameter as 0.01.

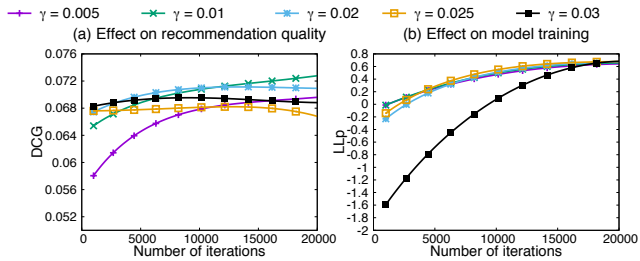

**Figure 3: Effect of the shrinkage parameter ($\gamma$).**

**Effect of different features.** BoostJet combines 250 different features (Trackers and Offer2Vec) using MatrixNet belonging to 5 different types. However, each of these different types of features has different impact on the learning task. We now evaluate the impact of these features on the recommendation quality. For each given feature type $T$, we consider all the other features except those of type $T$ to train the MatrixNet model and then provide the recommendations. This way we evaluate the impact of the feature type $T$ in the recommendation quality. Figure 4 demonstrates this impact from different types of features on the recommendation quality in terms of DCG. We see that the best quality is achieved by using all the features (BoostJet). We also observe that the temporal feature type has the highest impact on the recommendation quality followed by the consumption pattern one (Offer2Vec). This effect of these two types of features is intuitive also as it effectively captures the behavioral changes among users as well as the offer-to-offer relations.

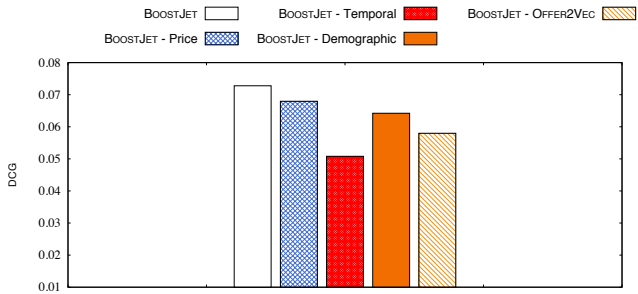

**Figure 4: Effect of different feature types in BoostJet. (Here "BoostJet- X" denotes the absence of features of type X.)**

**Feature strength analysis.** We next analyze the effect that we observed for different types of features. For this purpose, we use *feature importance* (effect) as our analytical measure. MatrixNet computes the relative importance of the features similar to [7]. We group the different features based on their types into 5 categories namely: *demographic*, *temporal*, *price*, *content* and *pattern*. Figure 5 compares the effect of these different features along with the variability among features from the same category. We observe that the effect of the features belonging to the same category can vary depending on how good the specific feature is. We also observe that the best feature is the one based on Offer2Vec which is slightly better than the one with maximum effect among the temporal Trackers. It is also important to note that we currently use a single feature for consumption pattern where as there can be different varieties of temporal features as we mentioned in § 2.1.

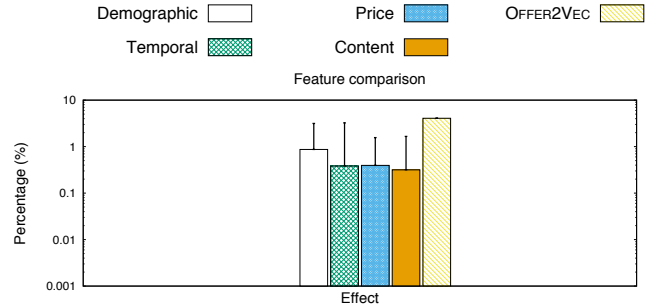

**Figure 5: Importance of different feature types in BoostJet.**

**Effect of personalized candidates.** We now analyze the effect of personalized candidates on which we use the classification task to generate the personalized recommendations. More precisely, in BoostJet, we have $20K$ popular offers from which we choose $5k$ best personalized candidates using the top-4 personalized Trackers along with Offer2Vec, and then the classification task is performed on these $5k$ personalized candidates. For the competitor, denoted by BoostJet-Pop, we perform the classification task on the $5k$ most popular candidates from the given shop. Note that the step for filtering the personalized candidates adds negligible overhead in terms of computation time since only 5 personalized features are used for the candidate generation compared to the 250 features used for recommendation. Table 4 demonstrates that the personalized candidate generation in BoostJet leads to around 13.1% improvement in recommendation quality in terms of DCG.

| Approach / Quality | BoostJet | BoostJet-Pop |
|---|---|---|
| DCG | 0.07278 | 0.06435 |

**Table 4: Effect of personalized candidates in BoostJet.**

**Comparison with implicit recommender.** We now compare the recommendation quality provided by BoostJet with another state-of-the-art technique leveraging matrix factorization for implicit feedback (i-Tencent) which is used in video recommendation service. i-Tencent treats the training data as indication of positive and negative preferences associated with significantly varying confidence levels. We observe from Table 5 that the quality provided by BoostJet is nearly 3 times better than the quality provided by i-Tencent which is attributed to the fact that BoostJet leverages different features concurrently. Hence, we see that BoostJet indeed achieves significant improvement in the recommendation quality.

| Approach / Quality | BoostJet | i-Tencent |
|---|---|---|
| DCG | 0.07278 | 0.02435 |

**Table 5: Quality comparison with state-of-the-art.**

## 4.3 Scalability

Given the huge volume of data and the different types of features (TRACKERS and OFFER2VEC) used in BOOSTJET, another crucial requirement is the scalability of the recommender. We now evaluate the capability of BOOSTJET to scale out on multiple machines. We measure the scalability in terms of the computation time as well as the speedup. The computation time in BOOSTJET involves the generation of features as well as training the MATRIXNET model whereas the speedup is measured in comparison to a minimum of 500 machines. We observe in Figure 6(a) that BOOSTJET achieves a linear speedup with an increasing number of computational resources. BOOSTJET finishes one complete recommendation round, generating recommendations for all the users in the test set, in 39 minutes with 1500 machines (Figure 6(b)).

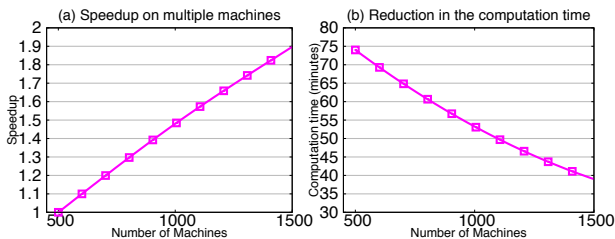

Figure 6: Scalability of BOOSTJET on multiple machines.

## 5 RELATED WORK

There has been significant work in recommender systems taking into account different recommendation features.

**Implicit vs explicit feedback.** There has been a lot of research on improving the recommendation quality given explicit feedback from users like ratings [1, 3]. However, explicit feedback is difficult to collect in online platforms as highlighted by Lee et al. [19]. There has been techniques developed for leveraging implicit feedback to provide recommendations. HOSLIM [4] computes higher order relations between items in consumed itemsets where the objective is to select the relations that maximize the recommendation quality. However, this approach does not scale for orders superior to pairs of items and hence cannot be used in practice. I-TENCENT, used in industry [14], treats training data as indication of positive and negative preferences associated with significantly varying confidence levels. We show that BOOSTJET significantly outperforms I-TENCENT.

**Temporal recommenders.** There has been some recent work on incorporating temporal dynamics in recommenders. Koren et al. [17] introduced a matrix factorization based model which identifies communal patterns of behaviour among users by modeling the way user and product characteristics change over time, in order to distill longer term trends from noisy patterns. Although widely popular, this model requires explicit ratings from the users to capture their behavioural drifts. However, as we mentioned earlier that explicit feedback is difficult to collect in online platforms whereas implicit feedback can be easily collected. In BOOSTJET, we leverage such implicit feedback to generate the TRACKERS which are used for recommendation computation.

**Revenue-driven recommenders.** Revenue is crucial to the operability of a recommender. As such there has been a few efforts in this direction, one of these is the generation of dynamic recommendations for maximizing the revenue [20] that takes into account a variety of factors including prices, valuations, saturation effects, and competition among different products. In BOOSTJET, similar factors are incorporated like financial capability of the users as well as the valuations of offers specific to a given shop while optimizing the learning task in MATRIXNET. Additionally, BOOSTJET also captures the changing behaviour of the user which is applicable to fluctuations in financial range as well.

**Demographic recommenders.** Demographic information is crucial in cold-start situations. Laila et al. [23] exploited demographic attributes of users for addressing cold-start problem in recommender. We leverage the demographic information primarily for cold-start users. However, the demographic information can indeed become more relevant for previous users when combined with other features like temporality or offer-content. For example, the information, that users from region A started purchasing offers of category B since last week, could indeed improve the recommendation quality by targeting offers from category B to users from region A. BOOSTJET enables this kind of recommendations which are not possible in standard demographic-based recommenders.

**Word-embedding based recommenders.** Recently, state-of-the-art results has been obtained, in the field of Natural Language Processing, on various linguistic tasks by learning latent representation of words using neural embedding algorithms. Barkan et al. [2] demonstrated that learning latent embeddings, using WORD2VEC, for items can enable the inferring of item-to-item relations even in the absence of user information. Similar approach is also used by Ozsoy et al. [22]. However, these models use WORD2VEC to learn the item embeddings which treat each user session separately and there is no connection between the context in different sessions. This problem is addressed in DOC2VEC where the session vectors are also input along with the item vectors for different sessions and hence the item vectors are linked across different sessions through the session vectors in the learning objective (Equation 1). BOOSTJET uses OFFER2VEC, our instantiation of DOC2VEC, to generate its features for consumption pattern for providing better recommendation quality.

**Feature combination for recommendation.** Feature combination is a recent research direction for recommenders. Recently, Covington et al. [5] demonstrated how different features can be combined with a deep neural network to provide video recommendations at YouTube. This approach leverages features like content (watch and search vectors), geographic embedding, age of a video. However, the context is significantly different between video recommendations and product recommendations based on features like price or temporal behaviour in shopping patterns. Hence, the feature selection (TRACKERS and OFFER2VEC in BOOSTJET) is crucial according to the recommendation objective.

## 6 CONCLUSION

In this paper, we show how we use TRACKERS to incorporate different recommendation features like price, temporality, demography, content as well as OFFER2VEC to incorporate users' consumption patterns. We combine these features from TRACKERS and OFFER2VEC in a novel recommender BOOSTJET which formulates the recommendation problem as a classification task. BOOSTJET uses gradient boosted machines to solve the classification task. Since the recommendation quality is dependent on the efficiency of the classification task, we supplement it with a negative sampling technique as well as a candidate generation technique to enhance the learning task. We show that BoostJet significantly improves the recommendation quality compared to another state-of-the-art approach leveraging standard implicit feedback. We also demonstrate that BoostJet scales linearly with an increasing number of computational resources. Lastly, it would be interesting to incorporate privacy [10, 11] while computing these statistical aggregates and thereby enabling privacy-aware recommendations.